\documentclass[twocolumn,showpacs,pra,floats]{revtex4}
\usepackage{amssymb}
\usepackage{graphicx}
\begin{document}
\setlength{\topmargin}{0.15in}
\setlength{\textheight}{8.25in}

\title{A graphical presentation of signal delays in the datasets of Weihs \emph{et al.}}
\author{Peter Morgan}
\email{peter.w.morgan@yale.edu}
\affiliation{Physics Department, Yale University, New Haven, CT 06520, USA.}
\homepage{http://pantheon.yale.edu/~PWM22}

\date{\today}
\begin{abstract}
A graphical presentation of the timing of avalanche photodiode events in the datasets from the
experiment of Weihs \emph{et al.} [\textit{Phys. Rev. Lett.} \textbf{81}, 5039 (1998)] makes manifest the
existence of two types of signal delay:
(1) The introduction of rapid switching of the input to a pair of transverse electro-optical modulators causes
a delay of approximately 20 nanoseconds for a proportion of coincident avalanche photodiode events;
this effect has been previously noted, but a different cause is suggested by the data as considered here.
(2) There are delays that depend on in which avalanche photodiode an event occurs;
this effect has also been previously noted even though it is only strongly apparent when the relative
time difference between avalanche photodiode events is near the stated $0.5$ nanosecond accuracy of
the timestamps (but it is identifiable because of 75 picosecond resolution).
The cause of the second effect is a difference between signal delays for the four avalanche photodiodes, for
which correction can be made by straightforward local adjustments (with almost no effect on the degree of violation
of Bell-CHSH inequalities).
\end{abstract}

\maketitle

\section{introduction}
There have been several analyses of the datasets generated by the experiment of Wiehs
\emph{et al.}~\cite{WeihsExpt,WeihsThesis}.
Two, due to Hnilo, Kovalsky, and Santiago~\cite{HKS}, and Adenier and Khrennikov~\cite{AKJPhysB2006}, are
particularly singled out by Weihs~\cite{WeihsComment}.
More recently, Zhao \emph{et al.} have analyzed the data as part of a program to construct a computational
model that violates Bell inequalities~\cite{Zhao}.

The Weihs \emph{et al.} datasets contain, for a number of runs, a recorded timestamp when an event occurred
in any of four avalanche photodiodes (APDs), together with a binary record of the input to a transverse
electro-optical modulator at the time of each APD event.
The timestamps have an accuracy of 0.5 nanoseconds and a resolution of 75 picoseconds.
The dataset for each run is in two locally recorded parts, labelled ``Alice'' and ``Bob'',
containing data collected at each end of the experiment.
The published papers and these datasets are the principal historical record of the Weihs \emph{et al.}
experiment, which together satisfy Bohr's requirement that an experiment must be classically described.
At this level of description, the datasets are essentially a nonlocal, contextual classical description of
the whole experiment's results, which are well-known to be incompatible with a local, noncontextual
classical particle model.

Section \ref{GP} describes a graphical presentation of three datasets, \emph{longdist35}, \emph{locbell2},
and \emph{bellstat1}, which differs from previous analysis by using the difference between APD timings as a
coordinate instead of considering coincidence of APD events as a binary value that is true if the
difference between their timings is less than the width of a chosen window (which in the case of Weihs
\emph{et al.}'s analysis was 4 nanoseconds, which the data suggests as an appropriate width).
The first of the three datasets considered uses two 500 meter fiber optic cables to distribute light from the
central light source and a Parametric Down-Conversion (PDC) crystal to two remote sites, at each of which
there is a transverse electro-optical modulator, the output of which is directed to a dynamically modified
polarizer and thence to a pair of APDs;
for the second and third of these datasets the transverse electro-optical modulators, the polarizers, and
the two pairs of APDs are close to the light source and the PDC crystal.
For the first two of these datasets, the input signals to the transverse electro-optical modulators are switched
rapidly between two values, whereas for the third dataset there is a constant input signal to the transverse
electro-optical modulators.

\begin{figure*}
\includegraphics{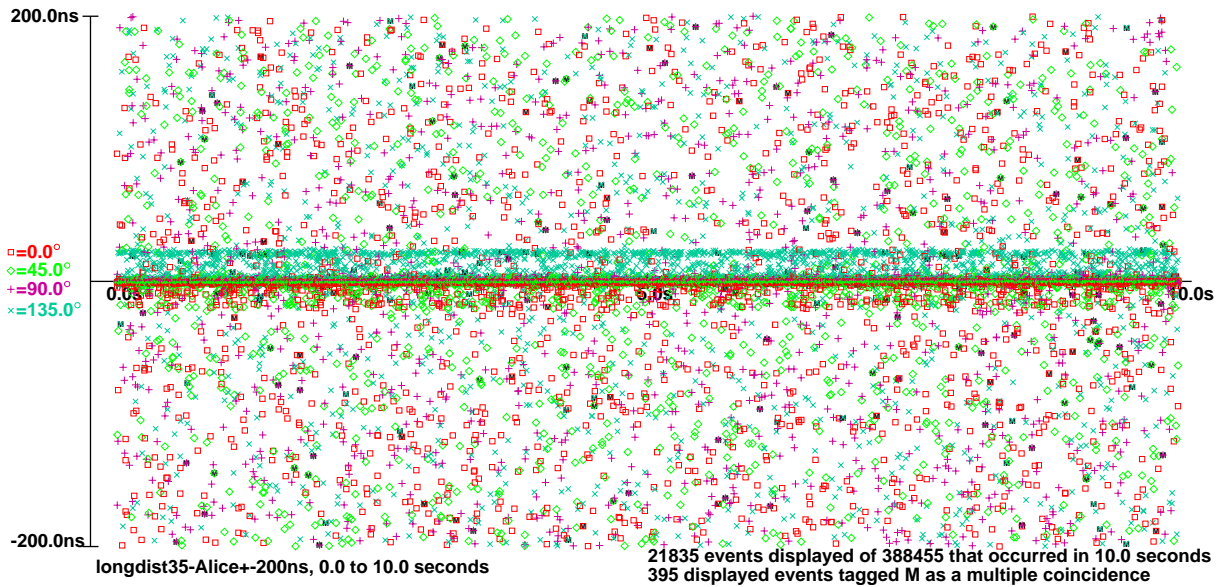}
\caption{\label{longdist35Alice200ns10s}
  (Color online) 10 second by $\pm 200$ nanosecond window on approximately coincident events from Alice's
perspective for \emph{longdist35}, with a single timing adjustment (described in Section \ref{GP}).}
\end{figure*}
\begin{figure*}
\includegraphics{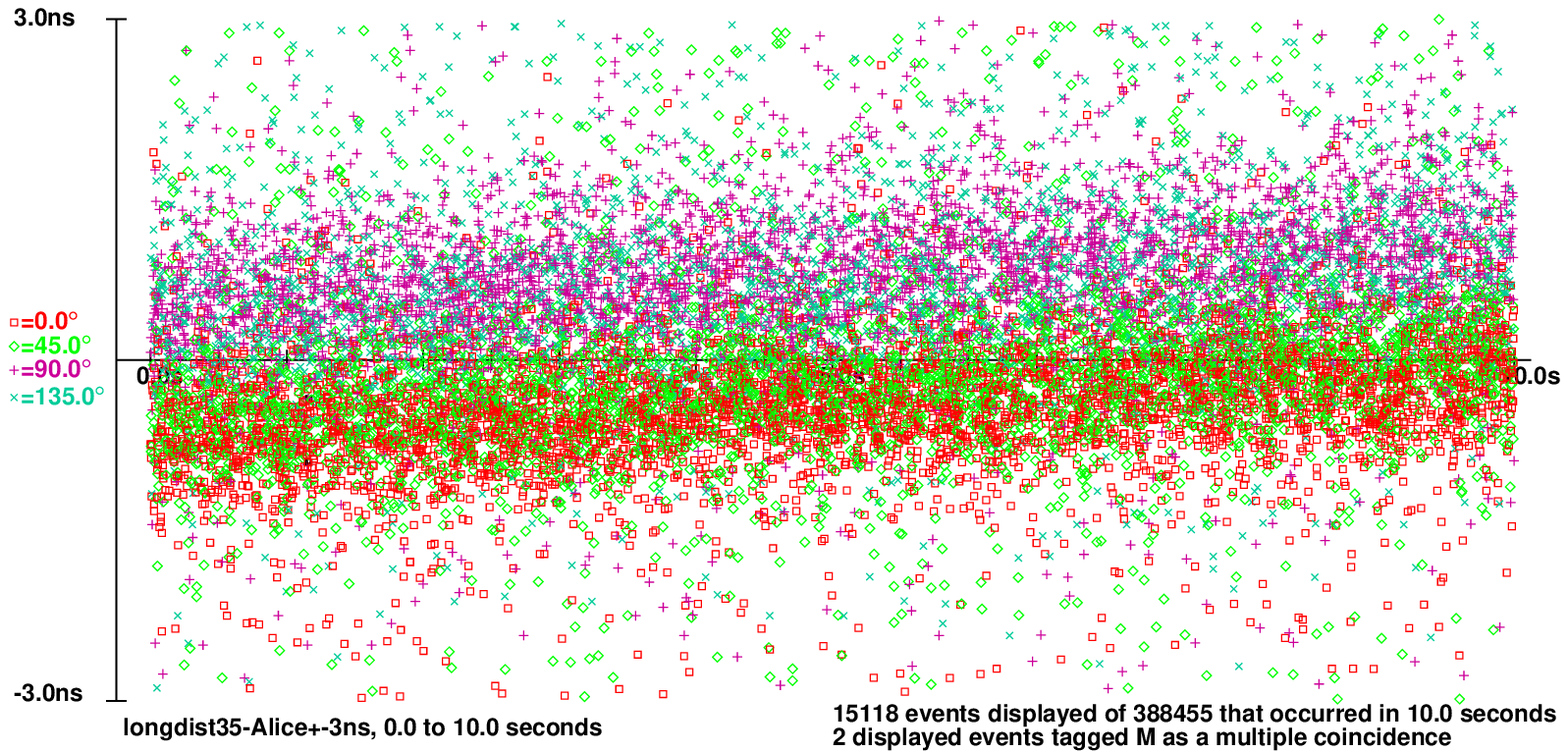}
\caption{\label{longdist35Alice3ns10s}
  (Color online) 10 second by $\pm 3$ nanosecond window on approximately coincident events from Alice's
perspective for \emph{longdist35}, with a single timing adjustment (described in Section \ref{GP}).}
\end{figure*}
\begin{figure*}
\includegraphics{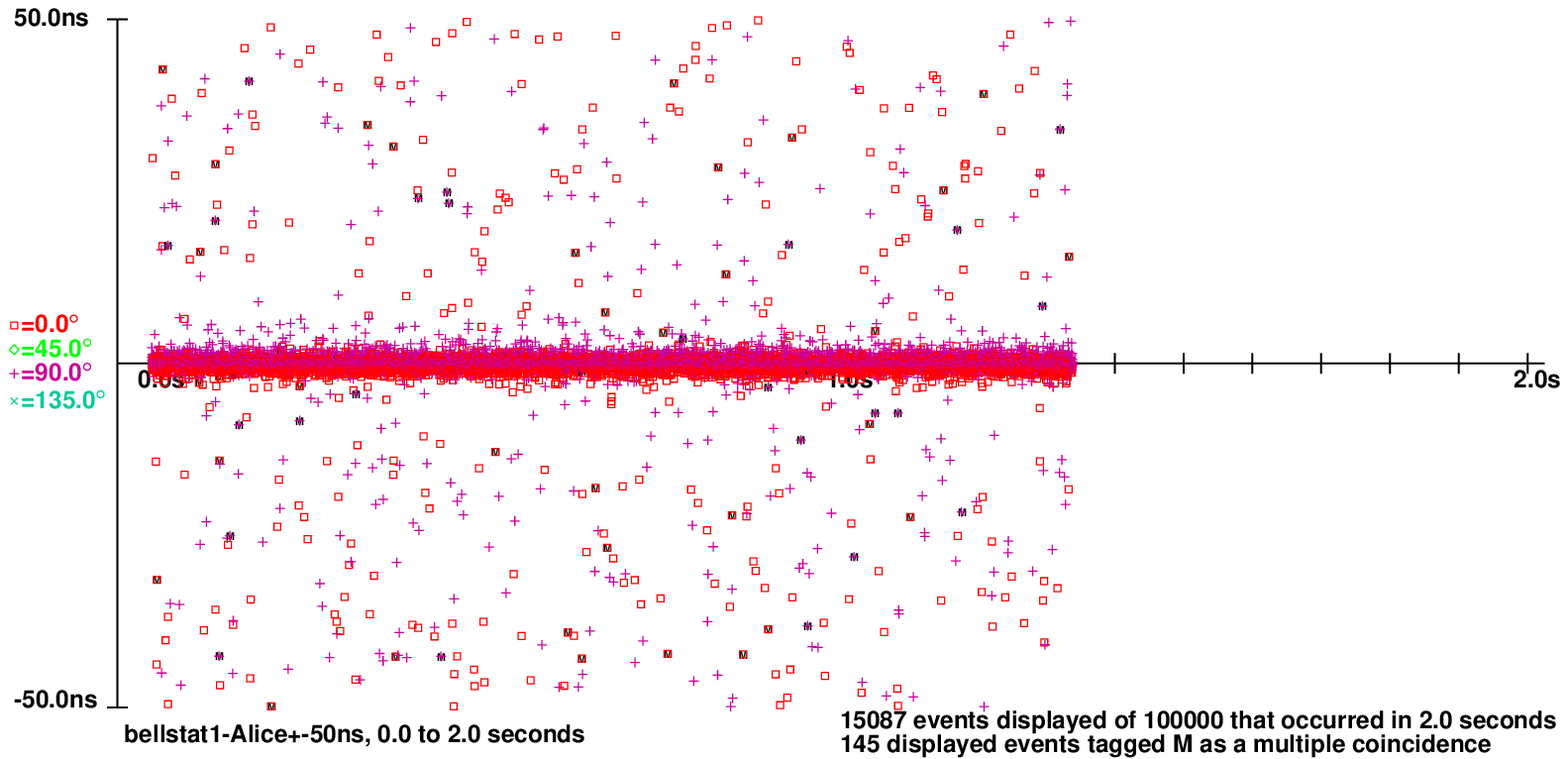}
\caption{\label{bellstat1Alice50ns2s}
  (Color online) 2 second by $\pm 50$ nanosecond window on approximately coincident events from Alice's
perspective for \emph{bellstat1}, with a single timing adjustment (described in Section \ref{GP}).}
\end{figure*}

The comparison of the first two datasets with the third dataset enables us to show in Section
\ref{EOMcorrelations} that rapid switching of the inputs to the transverse electro-optical modulators
causes a delay of approximately 20 nanoseconds for a proportion of coincident APD events.
We also show in Section \ref{longdist} that there is a correlation of APD events with the relative timing of APD events,
on a time-scale of approximately 1 nanosecond.
It is clear that these correlations are caused by different timing delays for the two APD signal channels (as
suggested to the author by Gregor Weihs), because adjusting the timings of the data for the two channels by a constant
offset eliminates the effect.
Both delays have been previously identified~\cite{AgueroEtAl}, however graphical presentation makes the delays
more immediately apparent.

Larsson and Gill~\cite{LG} show that there is a ``coincidence loophole'' that is distinct from the
``detection loophole'', which the Weihs \emph{et al.} experiment is well-known not to close.
The 1 nanosecond time-scale of the correlations described here, however, does not immediately
suggest a communication protocol that would allow a hidden variable model to be constructed.
In particular, the 1 nanosecond time-scale is less than the 4 nanosecond time-scale of typical window widths,
so these correlations are in a different regime than the timing correlations that are hypothetically
introduced by Zhao \emph{et al.}~\cite{Zhao} as part of a hidden variable model.

Although essentially self-contained, the analysis of the Weihs \emph{et al.} datasets that is described here
is partly motivated by a random field perspective on quantum
fields~\cite{MorganBellRandomFields,MorganRQKG,MorganLieFields}, which suggests a close look at the datasets
using methods of (stochastic) signal analysis.
In particular, precise differences of the relative timing of the coincidence of APD events seem more
likely to be of interest from a field perspective, instead of the less detailed binary fact of approximate
coincidence that is suggested by a quantized photon perspective.
Methods of signal analysis are also adopted by Hnilo, Kovalsky, and Santiago~\cite{HKS}.

\section{A graphical presentation of the datasets}\label{GP}
A C++ program was written to construct Embedded PostScript graphs for a given dataset, either from
Alice's perspective or from Bob's perspective.
For the case of Alice's perspective, we plot the time difference $t_A-(t_B(t_A)-\mathsf{offset})$ at each
time $t_A$ at which there is an APD event at $A$ (Alice's location), where $t_B(t_A)-\mathsf{offset}$ is
the closest time to $t_A$ at which there is an APD event at $B$ (Bob's location), with an offset that allows
for a slight difference between timestamps at $A$ and at $B$ that is more-or-less constant over time periods
of ten seconds (the offsets for the datasets considered here, \emph{longdist35}, \emph{locbell2}, and
\emph{bellstat1}, are $+4$ nanoseconds, $-12.5$ nanoseconds, and $+10.7$ nanoseconds, respectively, with
the sign reversed when considering Bob's perspective).
A negative value of $t_A-(t_B(t_A)-\mathsf{offset})$ corresponds to an APD event at $A$
being earlier than the APD event at $B$ that is closest in time to it.
At each coordinate $(t_A,t_A-(t_B(t_A)-\mathsf{offset}))$ a symbol is plotted that represents the binary
input to the transverse electro-optical modulator, which for the datasets discussed correspond either to
angles of $0^\circ$ or $45^\circ$ for Alice, or to angles of $22.5^\circ$ or $67.5^\circ$ for Bob.
The symbol plotted also represents in which of the two APDs there was an event, which is taken to add
$90^\circ$ to the angle.
Internally, the numbers $0$, $1$, $2$, and $3$ are recorded, which correspond, for the datasets discussed,
to angles $0^\circ$, $45^\circ$, $90^\circ$, and $135^\circ$ for Alice, and to angles $22.5^\circ$,
$67.5^\circ$, $112.5^\circ$, and $147.5^\circ$ for Bob, which are represented graphically by a square,
a diamond, a cross, and an ``$\mathsf{x}$''.
If the APD event at $A$ that is one of a coincident pair is close to more than one APD event at $B$, or if
the APD event at $B$ that is one of a coincident pair is close to more than one APD event at $A$, the
coincident pair is tagged as a multiple coincidence, represented graphically by an ``M'', and ignored for
the purposes of computing the violation of Bell-CHSH inequalities.
\begin{figure*}
\includegraphics{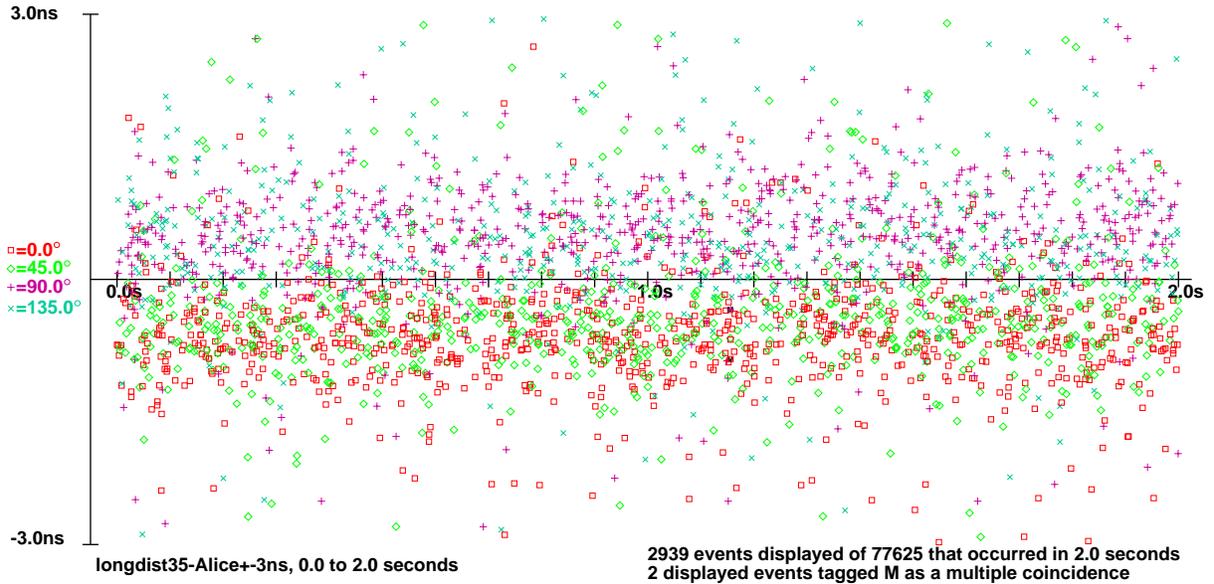}
\caption{\label{longdist35Alice3ns2s}
  (Color online) 2 second by $\pm 3$ nanosecond window on approximately coincident events from Alice's perspective
for \emph{longdist35}, with a single timing adjustment (described in Section \ref{GP}).
There is a visible asymmetry between which events are observed relatively early at $A$ and which events are observed
relatively late at $A$, compared to coincident events at $B$.}
\end{figure*}
\begin{figure*}
\includegraphics{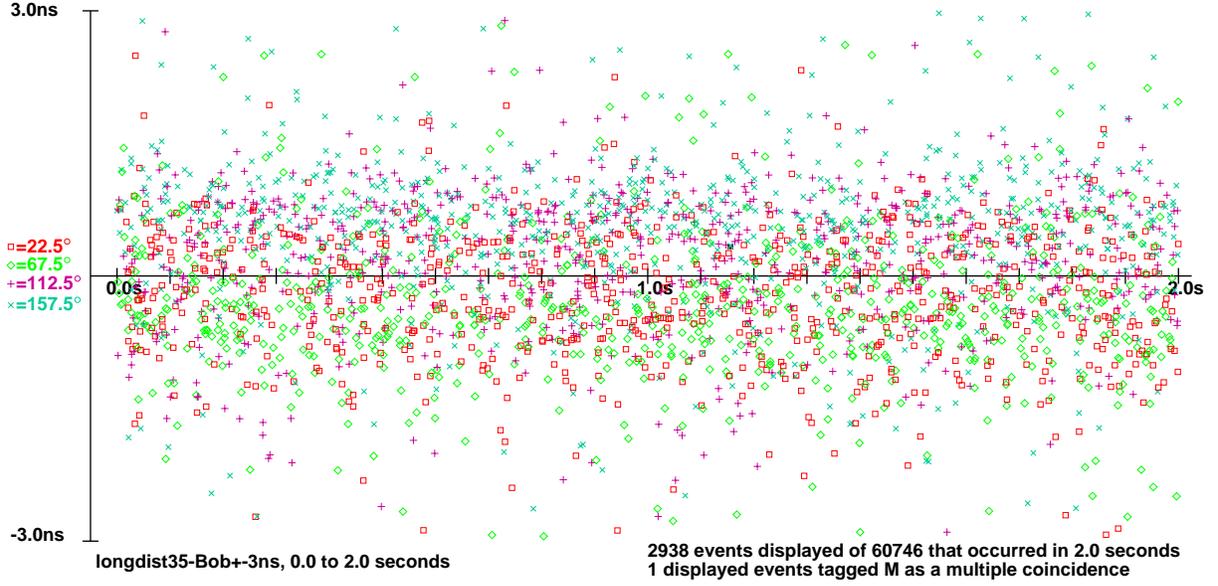}
\caption{\label{longdist35Bob3ns2s}
  (Color online) 2 second by $\pm 3$ nanosecond window on approximately coincident events from Bob's
perspective for \emph{longdist35}, with a single timing adjustment (described in Section \ref{GP}).
The temporal asymmetry of events is less apparent than in Figure \ref{longdist35Alice3ns2s}, but is still visible.}
\end{figure*}
\begin{figure*}
\includegraphics{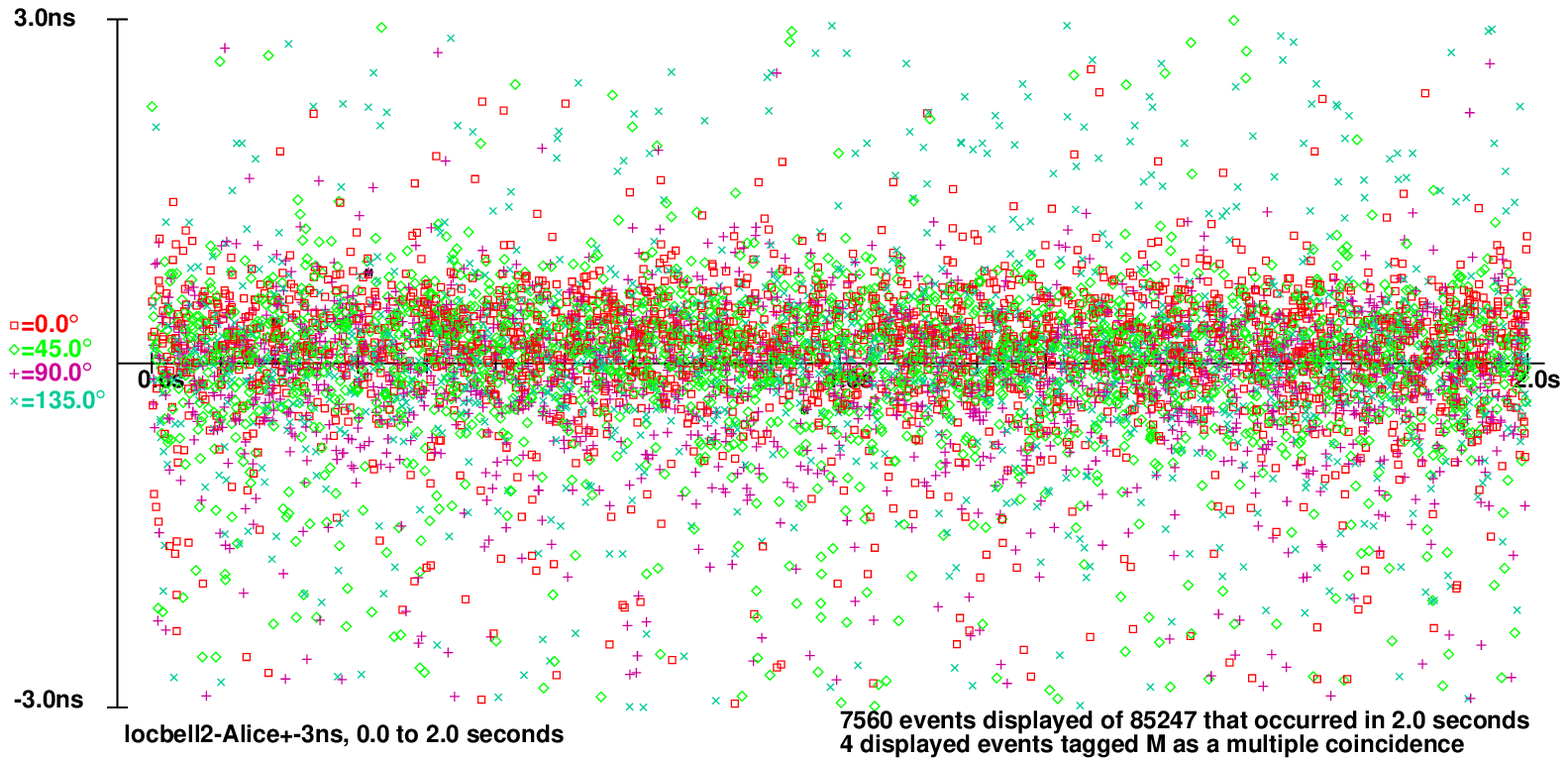}
\caption{\label{locbell2Alice3ns2s}
  (Color online) 2 second by $\pm 3$ nanosecond window on approximately coincident events from Alice's
perspective for \emph{locbell2}, with a single timing adjustment (described in Section \ref{GP}).
The density of closely coincident events is greater because the transverse electro-optical modulator and APDs
are not separated from the light source by 500 meters of fiber optic cable.
The temporal asymmetry of events visible in Figures \ref{longdist35Alice3ns2s} and \ref{longdist35Bob3ns2s}
is not visible here, but it can be made apparent using a graphical presentation such as
Figure \ref{longdist35Alice3nsHistogram10sWA}.}
\end{figure*}

\section{Delays caused by the transverse electro-optical modulator}\label{EOMcorrelations}
Figures \ref{longdist35Alice200ns10s} and \ref{longdist35Alice3ns10s} show a vertical $\pm 200$ nanosecond
section and a vertical $\pm 3$ nanosecond section of the graph for the whole ten seconds of data for
\emph{longdist35} from Alice's perspective; subsequently we will show graphs for only the first two seconds
of data, which are adequate for our purpose here, more readable, and result in smaller Embedded PostScript
files.
It is necessary to show different vertical sections of the graph to make different features of the data
visually apparent.
Note that there is a wide stripe within approximately $\pm 20$ nanoseconds in which there is a greater
density of APD events, and a narrower stripe within approximately $\pm 1$ nanosecond in which there is a
much greater density of APD events.
The density of APD events beyond $\pm 20$ nanoseconds is approximately uniform.
For comparison, the characteristic lengths of the elementary geometry of the experiment are a straight
line separation between $A$ and $B$ of 400 meters and a separation between $A$ and $B$ along the fiber optic
cable light guide of 1000 meters, corresponding to characteristic times of approximately 1.3 microseconds
and 3.3 microseconds at the speed of light.
In the vertical $\pm 3$ nanosecond section, an apparent drift of the relative times of APD
events at $A$ and $B$ over 10 seconds is visible, which is consistent with properties of the clocks that are
used to provide timestamps, described by Gregor Weihs as providing ``a relative accuracy of $8\cdot 10^{-11}$,
corresponding to $4.8$ [nanoseconds], for a $60$ [second] interval''\cite{WeihsComment}.
The more-or-less linear drift over a ten-second timescale allows for a straighforward adjustment of the timestamps.

The stripe of APD event pairs that are within $\pm 20$ nanoseconds of each other is also present in the
dataset \emph{locbell2} (not shown), from both Alice's and Bob's perspective, but it is not present in the
dataset \emph{bellstat1}, for which a vertical $\pm 50$ nanosecond section is shown in Figure
\ref{bellstat1Alice50ns2s}, nor in any of the datasets \emph{bellstat0}, \emph{bellstat2}, and \emph{bellstat3}.
The dataset \emph{bellstat1} is less than 2 seconds long; only two types of APD events
are present because there is no switching of the transverse electro-optical modulator.
Except for the dataset \emph{bellstat1}, the binary input to the transverse electro-optical modulator is
chosen randomly approximately every 100 nanoseconds.
It takes approximately 14 nanoseconds to switch from one setting to another, during which time APD events are
ignored~\cite{WeihsComment}.
Given the different distribution of APD event pairs when there is no rapid switching of the transverse
electro-optical modulator, and given that the 14 nanosecond period during which APD events are ignored is
a comparable time-scale to the 20 nanosecond broadening of coincidences of APD events, it seems likely
that the effect is caused by either electrical or optical properties of the rapid switching of the
transverse electro-optical modulator.
Ag\"uero \emph{et al.} suggest that the ``cause is a drift between the clocks at stations $A$ and $B$''
because their experiment uses a single clock and does not show a broadening of coincidences~\cite{AgueroEtAl},
however there is no reason to think that the clock is inaccurate on the relatively coarse scale of
$\pm 20$ nanoseconds in the Weihs \emph{et al.} experiment.

\section{Timing correlations in long distance datasets}\label{longdist}
\begin{figure*}
\includegraphics{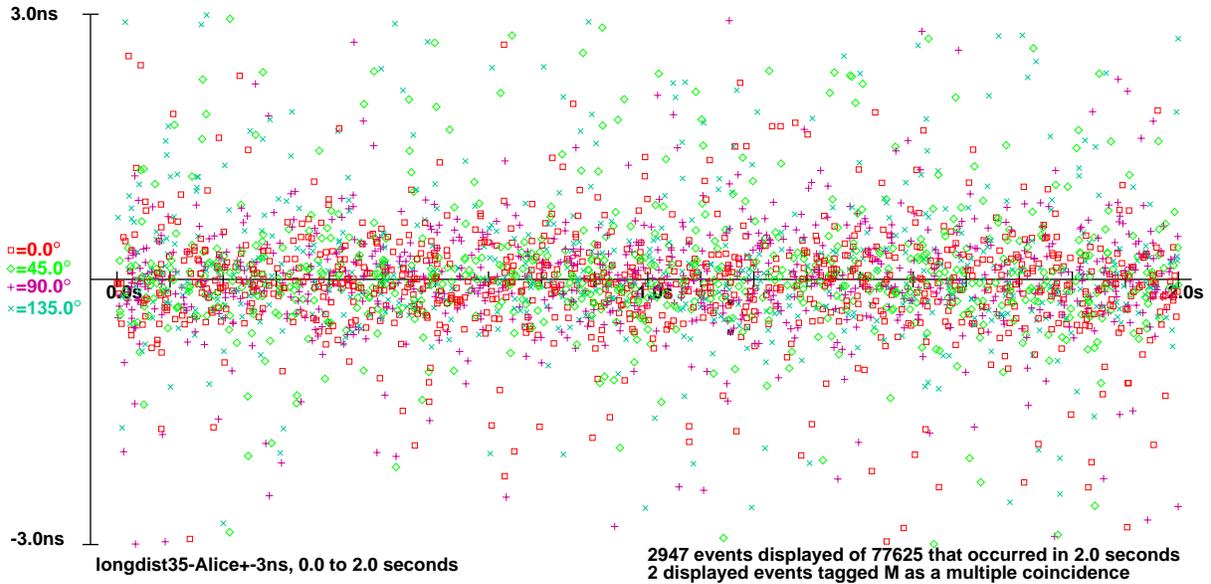}
\caption{\label{longdist35Alice3ns2sWA}
  (Color online) 2 second by $\pm 3$ nanosecond window on approximately coincident events from Alice's perspective
for \emph{longdist35}, adjusted for relative timing drift and for local detector delays.}
\end{figure*}
\begin{figure*}
\includegraphics{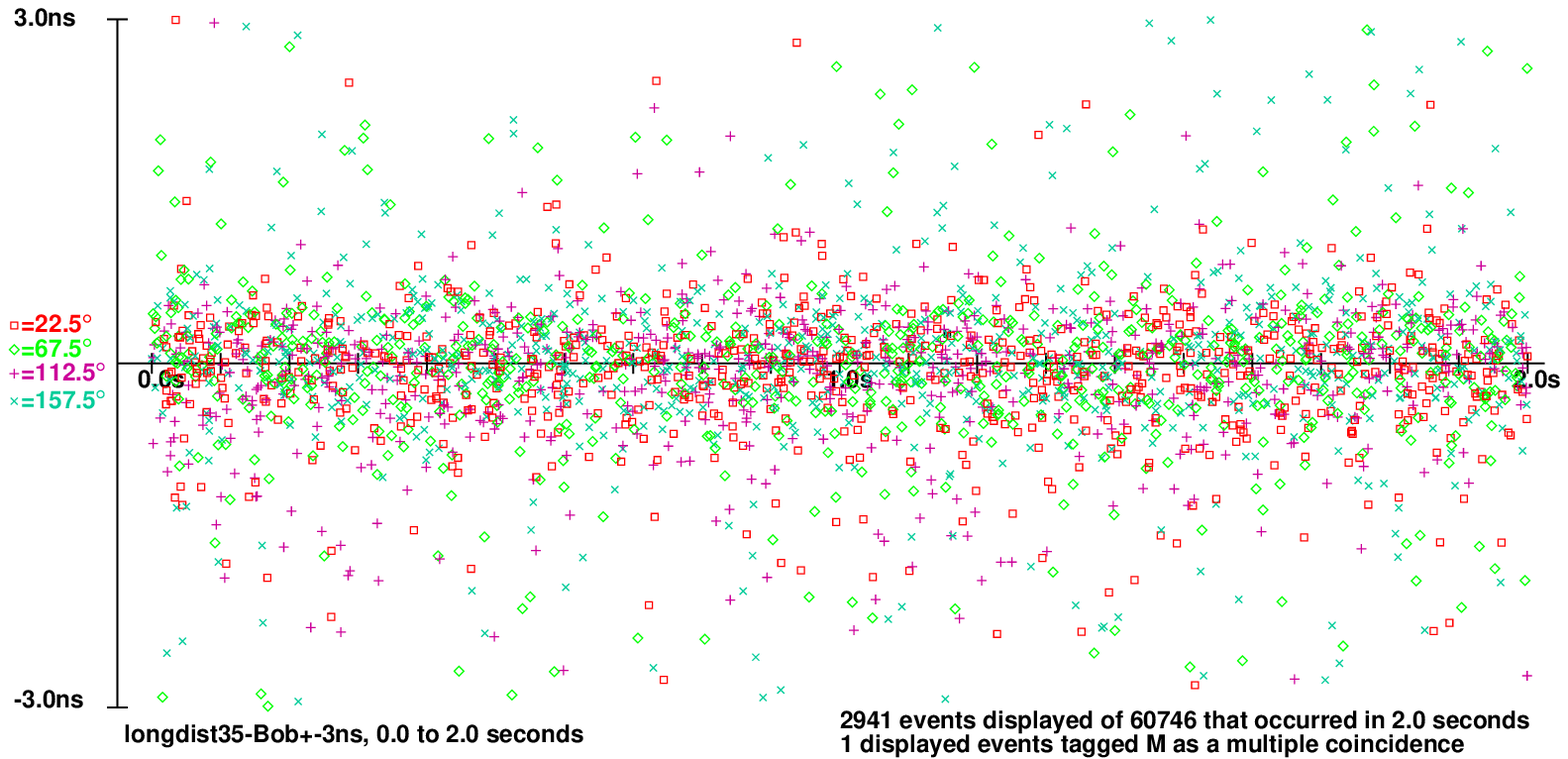}
\caption{\label{longdist35Bob3ns2sWA}
  (Color online) 2 second by $\pm 3$ nanosecond window on approximately coincident events from Bob's
perspective for \emph{longdist35}, adjusted for relative timing drift and for local detector delays.}
\end{figure*}
It is already apparent in Figure \ref{longdist35Alice3ns10s} that there is a strong correlation on a
nanosecond scale between the direction associated with APD events at $A$ and the time difference between
the APD events at $A$ and $B$.
The correlation is easily visible in Figure \ref{longdist35Alice3ns2s}, for the dataset \emph{longdist35}
(and is equally visible for the dataset \emph{longdist34}, which is not shown), but is not as evidently present
in Figure \ref{locbell2Alice3ns2s}, for the dataset \emph{locbell2}, the difference being the extensive
modification of the experimental apparatus necessary to introduce a much greater distance between the APDs and
two lengths of 500 meters of fiber optic cable for \emph{longdist35} (and for \emph{longdist34}).

The precision of the correlation is greater from Alice's perspective than from Bob's perspective, as can be
seen by comparison of Figures \ref{longdist35Alice3ns2s} and \ref{longdist35Bob3ns2s} (note that relative time
difference from Bob's perspective has the opposite sign to relative time difference from Alice's perspective).

The apparent correlation can be eliminated, however, by essentially local adjustments that depend only on which APD
the event occurred in, effectively because the electronics associated with the four APDs introduce slightly different
delays.
The consequent Figures \ref{longdist35Alice3ns2sWA} and \ref{longdist35Bob3ns2sWA} show the result of including both
an adjustment for timing drift of 50ps per second and local detector adjustments of 0.0ns and 0.9ns for Alice and
4.4ns and 4.7ns for Bob, instead of a uniform less detailed adjustment of 4 nanoseconds between Alice and Bob that is
used for Figures \ref{longdist35Alice3ns2s} and \ref{longdist35Bob3ns2s}.
The adjustments required for \emph{locbell2} are not as great.

\section{Conclusion}
Although the rapid switching of the transverse electro-optical modulator does not change the signal enough
to invalidate the results or interpretation of the Weihs \emph{et al.} experiment, nonetheless it is a
matter of concern that rapid switching introduces a distortion of APD event timings on a time-scale of
20 nanoseconds, which is enough to have a technological impact in some circumstances.
The local adjustments for different signal delays for different detectors are relatively small and of marginal
interest in the Weihs experiment.

\section{Discussion}
This paper is not intended for publication, so I include here fragmentary comments on quantum mechanical experiments
---for which Bell-violating experiments have now become the most common example, replacing the old favorite, the
double-slit experiment--- that are rather disconnected from the analysis of Gregor Weihs' experiment that
is the paper's main purpose, but that may be of interest to readers of the Bell inequality literature.

Although we can often describe experimental apparatus coarsely using operators and states that act on
low-dimensional Hilbert spaces, it is a nontrivial task to construct experimental apparatus that can be
described accurately in such simple terms.
If we wish to describe an experimental apparatus accurately we generally must resort to quantum field models,
unless we have taken great care to ensure that an experimental apparatus is dynamically elementary.
This is analogous to descriptions of a pendulum as a simple harmonic oscillator, which may be quite accurate
as a first approximation, even for something as complex as a child's swing, but it is no small thing to
construct a classical system that can be very accurately modeled as a simple harmonic oscillator, so that,
for example, it may be used as an accurate clock.
For the purposes of quantum computation and other technological applications of the mathematics of quantum
theory, it is desirable to construct devices that are accurately describable using operators and states that act
on low-dimensional Hilbert spaces, but it is a difficult engineering feat.

From a classical thermodynamics point of view, we expect the statistics of the APD events to be determined
by the geometrical configuration of the whole apparatus, because the configuration of the experimental
apparatus determines the properties of the coarse-grained equilibrium state of the electromagnetic field
and of the PDC crystal and APDs that interact with it.
Even though at a fine-grained level there are many non-equilibrium APD events, nonetheless in terms of
those statistics of the properties of the APD events that are generally reproducible on time-scales of
seconds or minutes, the experiment is at equilibrium.
Indeed, it is precisely required that at least some statistics must be reproducible in order for us
to claim that we are doing a Scientific experiment, but there is no rule about \emph{what}
statistics must be reproducible.
A classical thermodynamic point of view is therefore in line with Bohr's view that the configuration of the
whole experimental apparatus conditions the experimental results.
Furthermore, we observe discrete events only because we choose to design and use a thermodynamically
metastable system such as an APD that is capable of making transitions between thermodynamically distinguishable
states in different ways depending on its surrounding environment.
We choose to condition the output signal from an APD to be a binary value, and to record the times at which
the output signal makes a transition to its excited state, but at a more detailed level the output signal
can be modeled as a continuous current, which is ultimately more appropriate if we adopt a field theoretic
perspective to enable more detailed modeling of the APD as an integral part of the whole experimental apparatus.
Thermodynamic transitions of an APD should, if ideally constructed, occur randomly, at a minimal dark rate, if
there is no external signal, and statistics of dark rate APD events should exhibit as little statistically
significant structure as possible, so that when we introduce an external signal, statistics of the
APD events can be used to characterize the external signal as cleanly as possible.

\appendix
\section{An alternative visualization}
\begin{figure*}
\includegraphics{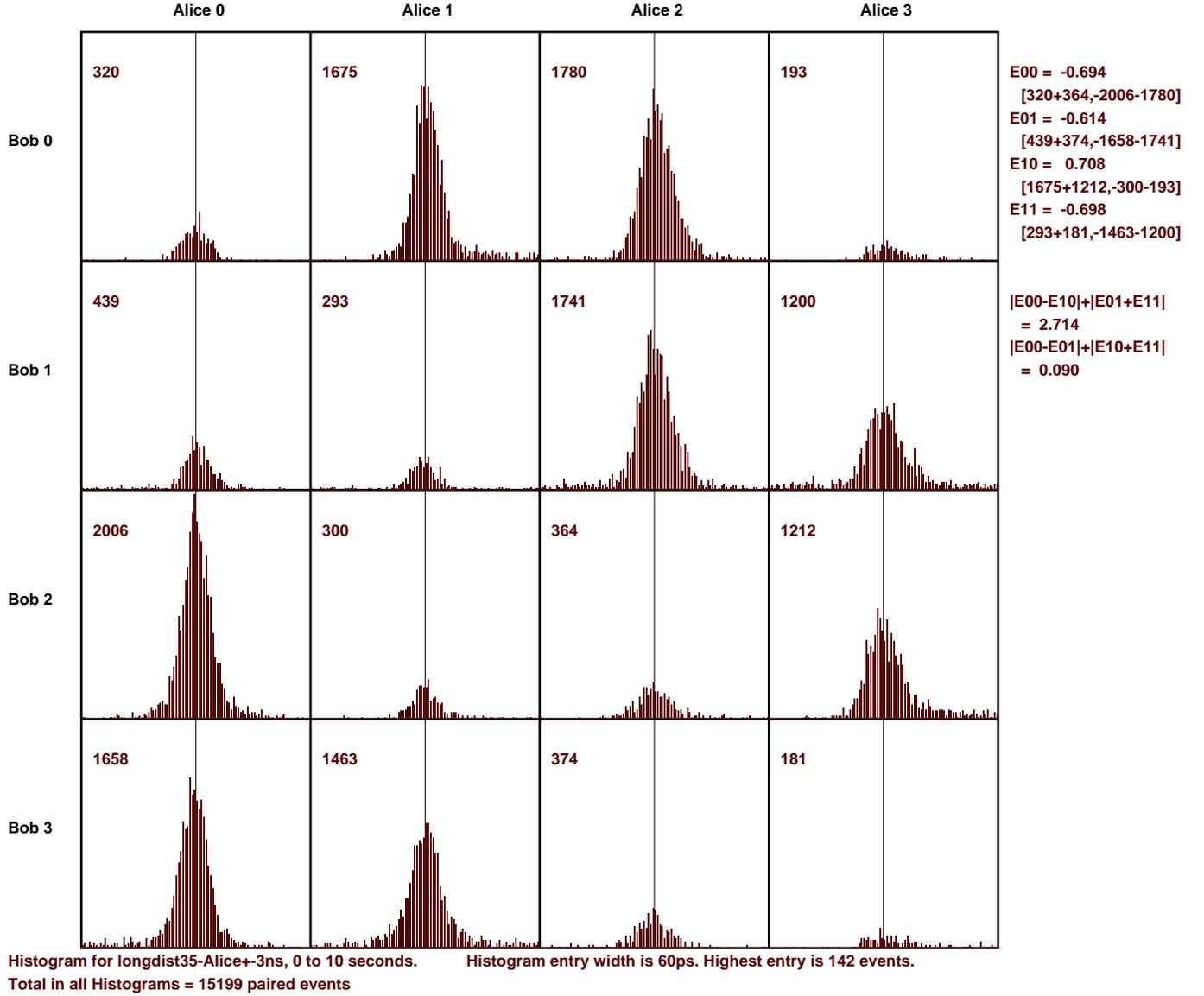}
\caption{\label{longdist35Alice3nsHistogram10sWA}
   Visualization of the detector and instrument setting timing statistics, for a 3ns window in 60ps sections, for
\emph{longdist35} from Alice's perspective, adjusted for relative timing drift and for local detector delays, together
with the corresponding calculation of the Bell-CHSH inequality violation.
For Alice, $0$, $1$, $2$, and $3$ correspond to angles $0^\circ$, $45^\circ$, $90^\circ$, and $135^\circ$.
For Bob, $0$, $1$, $2$, and $3$ correspond to angles $22.5^\circ$, $67.5^\circ$, $112.5^\circ$, and $147.5^\circ$.
}
\end{figure*}

Figure \ref{longdist35Alice3nsHistogram10sWA} presents a visualization of the detector and instrument setting
timing statistics, for a 3ns window in 60ps sections, for \emph{longdist35} from Alice's perspective, adjusted for relative
timing drift and for local detector delays, together with the corresponding calculation of the Bell-CHSH inequality violation,
for the complete 10 second run.
The vast majority of the detector coincidences are seen to lie within a $\pm$1ns window.
The derivation of the necessary local detector adjustments for the limited number of datasets considered was done by eye,
ensuring that as far as possible all 16 distributions are centered in this graphical presentation.

\begin{acknowledgments}
I am grateful to Gregor Weihs for giving me access to the datasets used above and for comments from him and from Alejandro Hnilo.
\end{acknowledgments}

\end{document}